# RHEED pattern classification by a convolutional neural network for the growth of chalcogenide thin films and nanostructures

Nathan Muetzel[1], Viet Luu[2], Sara Bey[1], Muhsin Abdul Karim[1], Kota Yoshimura[1], Xinyu Liu[1], Marwan Gebran[3], Badih A. Assaf[1,*]

1 Department of Physics and Astronomy, University of Notre Dame, Notre Dame, Indiana 46556, USA

2 QuarkNet Research Experience for High Schools, University of Notre Dame, Notre Dame, Indiana 46556, USA

3 Department of Chemistry and Physics, Saint Mary's College, Notre Dame, Indiana 46556, USA

*corresponding author: bassaf@nd.edu

**Abstract**

The use of reflection high energy electron diffraction (RHEED) plays a critical role for in-situ characterization in molecular beam epitaxy, pulsed laser deposition and sputtering. While sensitive to crystal symmetries and morphology, it is used ubiquitously to determine the growth modes of thin films. However, analysis of RHEED patterns depends on skilled experts and is therefore difficult to incorporate into the growth strategy in real-time. The development of machine learning (ML) processes, specifically convolutional neural networks (CNNs), presents a unique opportunity towards real-time RHEED pattern recognition. In this study, we develop a CNN model that can accurately classify four common and distinct RHEED patterns present in chalcogenide thin film growth. Its reached accuracy reached 94.9% for single run and 91.2% when averaged over 20 seeds. Our network is able to distinguish the nucleation of three common growth modes encountered in epitaxy, namely Volmer Weber, Stransky-Krastanov and Frank-van der Merwe, potentially enabling future automation of substrate temperature and shutter control informed by RHEED data. The network is material-agnostic and distinguishes the VW process with greater than 98% accuracy but is somewhat more limited in its ability to properly classify roughening and the initiation of Stransky-Krastanov growth. Our findings show that ML techniques can be successfully implemented even in cases where there is no detailed knowledge of growth chemistry providing an avenue towards real-time incorporation of ML to control nanostructure nucleation and thin film morphology.

**Introduction**

Reflection high energy electron diffraction (RHEED) has long played an integral role in thin film synthesis processes [1,2]. The use of RHEED gives growers real-time in-situ information on the development of various structures at the surface of materials. In molecular beam epitaxy (MBE), RHEED is commonly used to identify pristine Si and GaAs substrate surfaces before a material is synthesized on top [3]. It is also commonly used to monitor layer-by-layer nucleation, to identify reconstructed surfaces, and to monitor the morphology of thin films [4–6]. Over the past decade

RHEED was also made compatible with other synthesis tools such as atomic layer deposition [7]. It is also routinely used in pulsed laser deposition and sputtering.

RHEED is an extremely powerful and reliable tool when it comes to the identification of thin film morphologies. In epitaxy, morphology is determined by the competition between surface adhesion and adsorbate adhesion. Surface adhesion quantifies how likely it is for an adatom to attach to the substrate surface. Adsorbate adhesion determines how likely it is for an adatom to attach to a previously deposited adsorbate. When the former is strongest, atomic species nucleate at the surface of the film forming continuous smooth layers. This growth mode is referred to as the Frank van-der-Merwe mode (FM) [8,9]. When the latter dominates, atomic species nucleate more favorably on previously deposited material, creating nanometric islands. This is referred to as the Volmer-Weber (VW) growth mode and is routinely used to synthesize quantum dots of semiconductors [8,10–12]. An intermediate mode – the Stranski-Krastanov (SK) mode – yields both a continuous 2D layer with 3D islands on top [8,9]. The RHEED patterns resulting from the VW mode can be easily distinguished from the patterns of FM grown layers. The former exhibit spots due to reflection from crystalline facets of islands, while the latter yield continuous streaks. Conversely the SK mode can yield patterns that are typically referred to as modulated streaks [2].

Machine learning (ML) has emerged as a highly sought-after tool to process and classify data in quantum material science [13,14]. It has proven extremely useful in the context of diffraction data on crystals [15,16]. In many instances, the classification of diffraction maps is in its own respect important to carry out, without the need to quantitively analyze content. This is especially true for in-situ RHEED used in thin film growth. While previous work cited above has implemented the use of RHEED to automate decision making in MBE, these previous implementations were specifically optimized for a single material system. There is a need to test the reliability of ML tools when it comes to holistic image recognition in RHEED data sets that include "bad" data with growths that are run as optimization steps.

Convolutional neural networks (CNNs) can be used to isolate and extract important information from the image and classify the images into distinct categories [17]. Some progress has already been made in the classification of RHEED images, with high accuracy classifications being carried out for various GaAs surface reconstructions using their respective RHEED images [18]. Other advances have been made in attempting to integrate ML techniques in the growth process by identifying the point at which substrates are at an ideal temperature for InAs QD growth [19,20]. Generally speaking, QD synthesis can follow an established synthesis scheme, where streaky RHEED from the substrate surface signals that the growth can be initiated, and transformation to spotty RHEED signals the nucleation of quantum dots (see Fig. 1). Fig. 1 shows an example workflow that summarizes this specific process. ML models were also used to distinguish between growth modes of 2D transition metal dichalcogenide thin films [21]. On the fly RHEED analysis was also applied to oxide thin films [22]. Machine learning applied to RHEED enabled the recognition of various structures of $Fe_xO_y$ in-situ [16].

No prior work has, however, broadly investigated applying machine learning to the synthesis of mono-chalcogenide crystals. Crucially, while previous work has implemented the use of RHEED to automate decision making in MBE, these previous implementations were specifically optimized for a single material system. There is a need to test the reliability of ML tools when it comes to material-agnostic image recognition in RHEED data sets that include "bad" data with growths that are run as optimization steps. This is especially the case for qualitative ML-based analysis that only classifies images based on the nature of pattern, without analyzing the diffraction pattern quantitatively. Classification schemes are also less computationally extensive and potentially advantageous to use once their performance is tested and demonstrated.

Here, we have built a material-agnostic CNN trained on a comprehensive image data set that includes RHEED patterns acquired from the epitaxial growth of IV-VI thin films and quantum dots, zincblende CdTe films, as well as hexagonal and zincblende MnTe thin films. The dataset includes images from various growth modes (VW, FM and SK) that exhibit sharp streaks from flat films, sharp spots characteristic of QD facets, and diffuse and spotty streaks referred to as "modulated streaks" from rough films. Those three categories are described in ref. [2]. Examples are shown in Figure 2(a-d). We also classify RHEED exhibiting irregular spots from failed non-epitaxial quantum dot growths as the category "anomalous spots" (see Fig. 2(c)). This last category classifies spots appearing at reciprocal-space positions that cannot be explained by diffraction from the expected side facets of well-defined three-dimensional quantum dots. The CNN is tested on images taken on materials that appear in the training and one new material that did not appear in the training set.

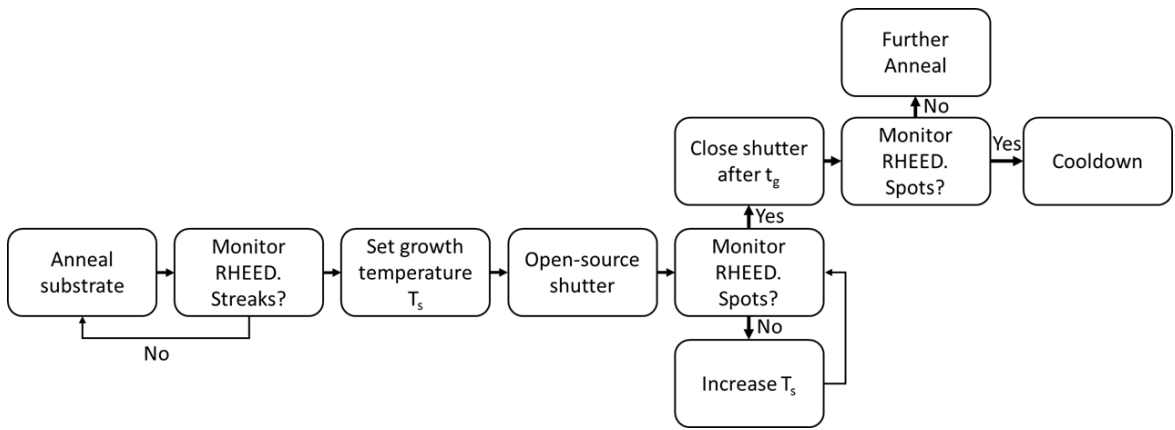

**FIG 1**. Example workflow for MBE synthesis of chalcogenide quantum dots.

Our approach focuses on the use of CNNs to identify and classify specific RHEED patterns resulting from various growth modes of these three-dimensional chalcogenides to ultimately automate growth optimization in real-time. We have thus implemented a highly accurate CNN – trained on this holistic dataset – which successfully identifies these patterns. The CNN is found to be capable of identifying the nucleation of quantum dots (VW growth) and their differentiation from 2D layers (FM growth) with great accuracy exceeding 98% for an optimized network. We also found that "modulated streak" patterns in RHEED resulting from SK growth or failed VW growth were classified with a lower accuracy. The reduction in accuracy was likely related to the fact that early stages of SK growth yielding anomalous streaks can sometimes be difficult to distinguish from FM growth, even by human users leading to challenges in labelling. The overall accuracy of the algorithm reaches 91.2%. The paper describes this algorithm, highlighting the ability of CNNs to predict growth outcomes from RHEED with an accuracy that is comparable to an experienced MBE expert but with instantaneous feedback enabling automated decision-making in MBE.

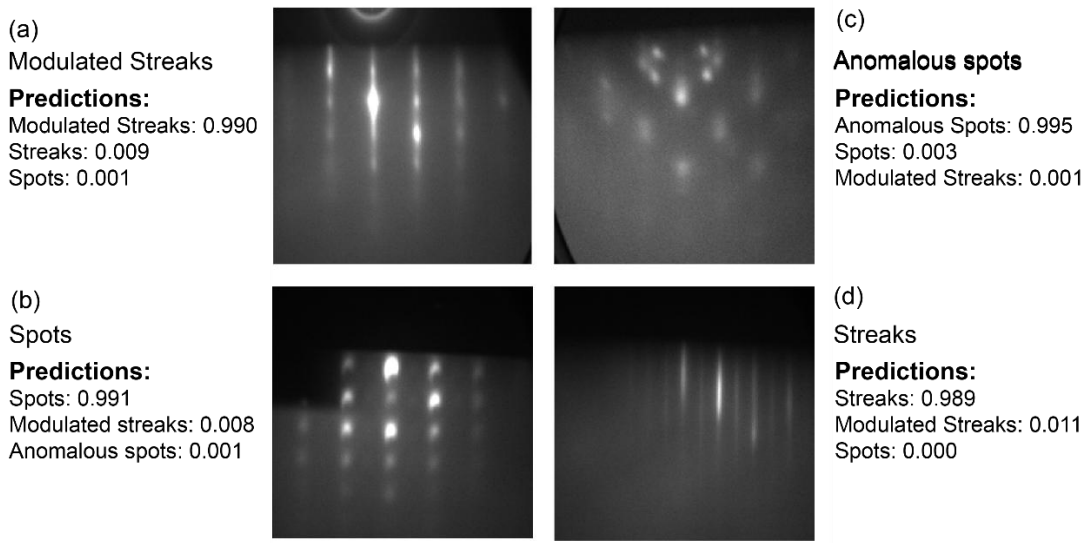

**FIG 2**. Sample images from each class that the model successfully predicted, along with the top three prediction probabilities of the model.

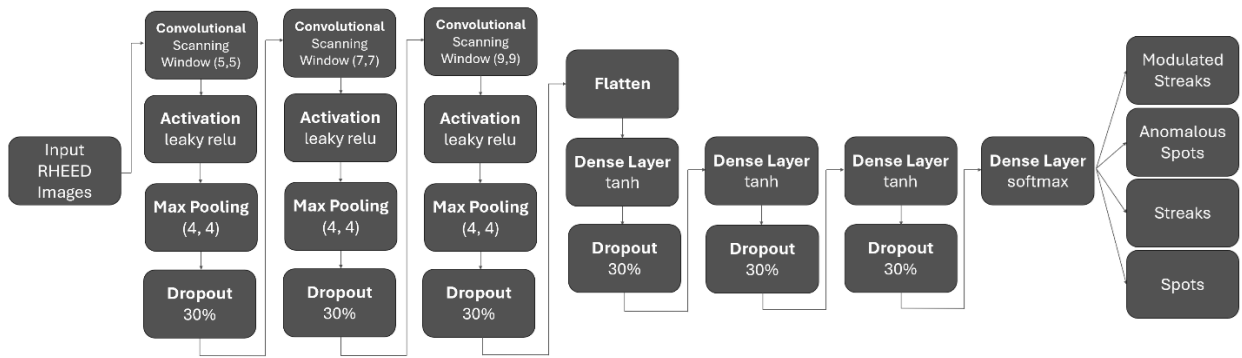

**FIG 3**. Diagram depicting the exact structure of the most successful neural network, trained and tested on 224 × 224 pixel images.

## Methods

The RHEED images used were obtained from growth of the following materials over a period of one year between July 2024 and July 2025: (Pb,Sn)Se quantum dots and films grown on BaF$_2$(111), (Pb,Sn)Te quantum dots grown on a CdTe(100) buffer, as well as CdTe(100), grown on GaAs(100) GeTe, (Ge,Mn)Te and (Ge,Sn)Te films grown on BaF2(111) and MnTe films grown on GaAs(111) and SrF$_2$(111) (see sample list and RHEED image examples for each material type in the supplementary material) [6,23–25]. The image library consisted of 809 raw images taken at different instances of the growths. The images were manually labeled and sorted into 4 categories: spots, streaks, modulated streaks, and anomalous spots. Figure 2 shows an example of each. Data augmentation was then applied to the images so that the number of images in each category would match the number of images in the most robust category, which was streaks. Flips about the vertical and horizontal direction as well as changes to the contrast, brightness, and saturation of the images were applied to create the artificial images, which brought the total number to 1,560. The balanced dataset was then fed into the CNN with a 60%/20%/20% split for train, validation, and test data respectively. The raw images were 780 × 580 pixels but were resized down and tested at sizes from 224 × 224 to 64 × 64 to save computing power and maximize efficiency of the network. Additionally, each model was evaluated based on its average performance on three different random seeds, or random mixings of data, to judge its ability to perform on all datasets.

A standard CNN features two distinct sections, one housing the convolutional layers (layers that apply a convolution operation) and one the dense layers (layers that are entirely connected to previous layers). The pixel content of the image is fed into the convolutional layer as a 2D array. Within the convolutional layer, multiple convolutional filters (kernels) convolve over the image and isolate the most important features. Once complete, the data is flattened into a 1D array and passed through the dense layers, which function as a traditional neural network that adjusts biases between nodes. The dense layers project to four logits which are converted to the categorial predictions via a softmax function. Because of the complexity of these networks, finding the most successful network requires manual tuning of the hyperparameters. These hyperparameters include the number of layers (3,3) and (3,2) and nodes in both the convolutional and dense sections, the activation functions, the optimizer, the learning rates (0.001, 0.0001 and scheduler), the image dimensions and the size of the scanning window. The scanning window represents the number of pixels that are evaluated simultaneously within the convolutional layers and condensed into one number for the next layer. To identify the most successful network, combinations of these hyperparameters were tested and the resulting train, validation, and test accuracies were reported. The optimization is carried out using an algorithm similar to what is reported in [26,27]. The most successful model, shown in Figure 3, consisted of three convolutional layers and three dense layers with dropout after each layer to combat against overfitting. Further tuning was done based on these results with additional information being obtained from figures such as plots of the loss function and accuracy vs epoch and confusion matrices. The hyperparameter space is discussed in the more detail in the supplement.

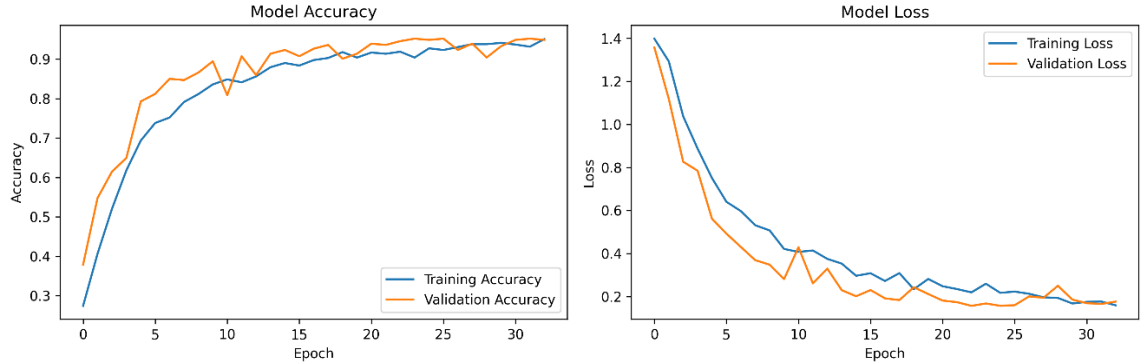

**FIG 4.** Training and Validation accuracy (left) and loss (right) plots vs epoch for a single run of the most successful model.

**Results and Discussion**

Figure 4 displays the improvements in both the accuracy and loss of the model as the number of epochs increases. For this single run, the accuracy of the model increases drastically initially before ultimately reaching a training accuracy of nearly 99% and a validation accuracy of nearly 95%. Similarly, the loss function significantly decreases before leveling out a minimum at higher epochs. This demonstrates the successful learning ability of the model to reach an apparent maximum in accuracy and minimum in loss. We carefully study the accuracy of the model and its ability to distinguish the four classes identified above.

We produce a confusion matrix of the real labels of the images versus the predicted label, shown in Figure 5. The analysis is carried out on the test dataset that contains 79 images in each category. The confusion matrix in Fig. 5(a) shows data classified over 20 distinct seeds, obtained by averaging the performance over the 20 seeds in order expose the network to different arrangements of the training, validation and test data. The average accuracy over these 20 seeds was 91.2% ± 1.8%, 91.5% ± 1.6%, and 98.8% ± 0.5% for test, validation, and training respectively. The matrix shows the model's ability to classify spots and normal streaks. These two classes are also well-distinguished from each other and are confused with each other less than 0.5% of the time. This is true across all runs, with a general performance of >99% on just distinguishing those categories. The four bottom right cells in the table reveal this.

Anomalous spots are well distinguished from streak categories but are sometimes confused with spots. Overall anomalous spots are identified with a high accuracy of 94.2%. The model is still able to accurately classify most modulated streaks but with an accuracy of nearly (85±4)%. Modulated streaks are often confused with normal streaks and normal spots. We discuss the origin of this confusion later in the next section. The matrix in Fig. 5(b) shows the confusion matrix for a single best run. The total accuracy of this single instance of the model is as high as 94.9% on test data, 94.9% on validation data, and 98.8% on training data. As for Fig. 5(a), modulated streaks and streaks are the most challenging to distinguish.

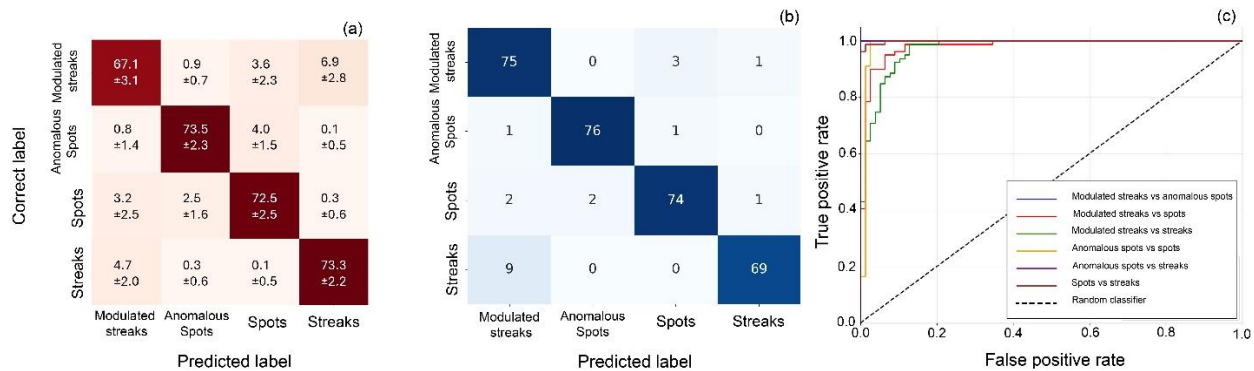

**FIG 5**. (a) Confusion matrix of the general performance of the best model averaged over 20 distinct seeds. (b) Confusion Matrix of a run of the most successful model, highlighting the ability to successfully distinguish between streaks and spots. Shown in each box are the average number of images falling into each category plus the standard deviation (taken over 20 seeds in (a)). There are 79 images in each category of the test dataset. (c) Receiver operating characteristic analysis.

We next plot a receiver operating characteristic (ROC) curve for one seed as an additional metric quantifying the performance of our classification scheme (Fig. 5(c)). This curve displays the corresponding true positive and false positive rates for every possible threshold of classification. A threshold can be thought of as a lower bound for the level of confidence needed to classify an image in a category. We then extract the area under the curve (AUC) which can be interpreted as follows. An AUC of 1.0 corresponds to the model correctly classifying all images at all thresholds, whereas an AUC of 0.5 corresponds to random guessing (dashed line in Fig. 5(c)). The area under the ROC thus quantifies the precision of a classification scheme. The extracted area under each ROC is shown in table I. The findings are consistent with the confusion matrix, yielding the lowest area for "modulated streaks" versus "streaks" at 0.973. Overall, all classes are classified with a ROC larger than 0.97.

|  | Area under the ROC curve |
|---|---|
| Spot versus streak | 0.999 |
| Anomalous spots versus streaks | 1 |
| Modulated streaks versus anomalous spots | 1 |
| Anomalous spots versus spots | 0.988 |
| Modulated streaks versus spots | 0.981 |
| Modulated streaks versus streak | 0.973 |

Table I. Area under the curve extracted from Fig. 5(c).

It is useful to discuss difficulties in identifying modulated streaks in RHEED, which are characteristic of rough, stepped surfaces. At a raw image level, these images contain a "mixture" of streaky and spotty patterns. Because of this, there is not always a concrete marker as to what fully classifies an image as having modulation in the streaky pattern at this level. Rather, what we have is a continuous distribution of growth patterns that must be cut somewhere for the sake of classification. Therefore, a highly accurate classification of these images seems to be extremely unlikely without a more quantitative description of the modulation pattern.

When a CNN predicts the class of an image, its output is a vector of the probabilities of the image falling into each of the classes. This means that an image is generally not classified with 100% certainty but rather is classified into the category with the highest probability of being correct according to the model. Figure 1 demonstrates an example of four images predicted correctly by the most successful model. The model is extremely confident in these classifications, with nearly 99% confidence in all instances. Figure 5 shows the same metrics for the same model but instead for images that were incorrectly classified. Here, the prediction probabilities greatly decrease, especially in the case of confusion involving modulated streaks. Each image gives a case example of an issue that we encountered.

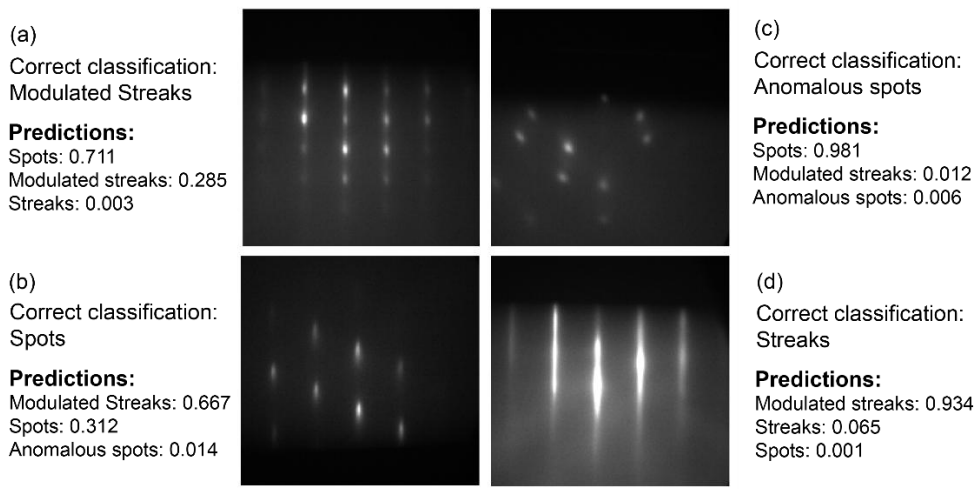

**FIG 6**. Sample images that the model incorrectly classified.

Fig. 6(a,b) are interesting situations where the network's accuracy is likely being limited by the labeling process of the training data. The probabilities are closer to 70% for the highest category and 30% for the second highest, demonstrating a low level of confidence. In both cases spots are visible but are either elongated or connected by faint streaks. A user labeled Fig. 6(a) as modulated streaks, while the algorithm favors the more visible spots as the main classifying signature. In Fig. 5(b), the opposite occurs, and while the user labels the figure as spotty (with elongated spots), the algorithm favors classification under "modulated streaks" but with a low certainty. This type of misclassification likely requires a more quantitative definition of what a modulate streak to sort images. Quantitative parameters can include the period of the modulation, the peak-to-peak amplitude of RHEED streak across the modulation or the spacing between streaks over the length of the streak.

Fig. 6(c) is a less interesting mistake, that can usually be overcome with improved data acquisition. It is an anomalous spotty pattern obtained while the sample is rotating. It is captured with spots somewhat off-center yielding difficulties in identifying their correct positions. The network fails to properly sort it in the "anomalous spots" category and instead classifies it under "spots".

Fig. 6(d), is an image showing streaks, with a weak modulation of their intensity due to the presence of the reflected direct beam near the central streak. The model classifies this image under modulated streaks, whereas in reality the modulation is artificial. This type of error is difficult to correct because it is often dependent on hardware settings and their state. Its likelihood of occurring should decrease with training on even larger datasets where similar images occur more frequently and are properly classified. A quantitative scheme for classification that includes sorting the diffraction streaks by separation and shape could also reduce the likelihood of misclassification here. It is however important to recognize that the overall success rate of the classifier is still very good, and reaches 91.2% in the testing phase.

We lastly test CNN's performance on a set of seven RHEED images acquired during the growth of $NbSe_2$, a material that did not appear in the training set. The model correctly classifies those seven images. The results are included in supplementary section 3.

**Conclusion**

The development of a material-agnostic CNN that can successfully classify RHEED patterns marks a significant step forward in the automation of chalcogenide thin film and QD growth. Our findings demonstrate that a model can achieve a 91.2% accuracy on all RHEED images deriving from a variety of different growth modes. We also found an upper limit to the accuracy of a model when classifying images that span across growth modes. This upper limit coincides with the abilities of human users to correctly and consistently label images, leading to a model that classifies images at nearly the same level. Additionally, the model will only become more accurate as the number of available RHEED images in the dataset increases. We also observe that "modulated streaks" in RHEED resulting from rough surfaces in the SK growth mode are not simple to distinguish from "spots" in the VW mode and normal "streak" in the FM mode. The control of substrate temperature in MBE for example can heavily impact whether and when surface roughening starts to develop [5,28]. Roughening is also initiated at the onset of strain relaxation with increasing thickness [29,30]. RHEED-informed automation of temperature control or shutter control will thus fail to recognize roughening nearly 10% of the time (in 7 out of 78 cases in Fig. 5(a)), until it is significant. A physics-informed network that can quantitatively interpret the pattern may improve performance. Despite its simplicity, the algorithm is, however, capable of reliably making binary decisions to identify the FM and VW growth modes. This fact is consistent with prior demonstrated successes in the implementation of RHEED analysis into automated MBE growth processes [19,20]. The development of our model and the study of its limitations allow for future real-time implementation of automation into the growth process. Future improvements can include training the network on images representing different surface reconstructions of common substrates, to automate substrate preparation more accurately.

**Supplementary material.** See supplementary material for full sample list, full list of CNN hyperparameters and additional testing using NbSe$_2$ RHEED images.

**Acknowledgement.** NM, XL, BAA and SB are supported by NSF-DMR-2313441. The synthesis and characterization of MnTe and MnTe-related films was supported by NSF-DMR-2313441. KY was supported by DE-SC0024291 for the synthesis and characterization of (Sn,Ge,In)Te and NbSe$_2$. This material (QD synthesis) is based upon work by MAK supported by the Center for Quantum Technologies under the Industry-University Cooperative Research Center Program at the US National Science Foundation under Grant No 2224985. VL acknowledges support from the Notre Quarknet Center.

The data that support the findings of this study are openly available at https://doi.org/10.5281/zenodo.17476777.